\documentclass[12pt,twocolumn]{iopart}
\usepackage[pdftex]{graphicx}
\begin{document}
\title[Effect of resistivity on silicon porosification...]
{Effect of silicon resistivity on its porosification using metal induced chemical etching}
\author{Shailendra K Saxena$^1$, Gayatri Sahu$^1$, P. K. Sahoo$^2$, Pankaj R Sagdeo$^{1,3}$ and Rajesh Kumar$^{1,3}$ \footnote{Corresponding Author: rajeshkumar@iiti.ac.in}}
\address{$^1$Material Research Laboratory, Discipline of Physics, School of Basic Sciences, Indian Institute of Technology Indore, Madhya Pradesh-452017, India}
\address{$^2$ School of Physical sciences, National Institute of Science Education and Research Bhubaneswar, Sainik School-751005, Odisha, India}
\address{$^3$ Material Science and Engineering Group, Indian Institute of Technology Indore, Madhya Pradesh-452017, India}
\begin{abstract}
  A comparison of porous structures formed from silicon (Si) wafers with different resistivities has been reported here based on the morphological studies carried out using scanning electron microscope (SEM). The porous Si samples have been prepared using metal induced etching (MIE) technique from two different Si wafers having two different resistivities. It is observed that porous Si containing well aligned Si nanowires are formed from high resistivity (1-20 $\Omega$cm) Si wafer whereas interconnected pores or cheese like structures are formed from low resistivity (0.02 $\Omega$cm ) Si wafers after MIE.  An explanation for the different porosification processes has also been proposed based on the initial doping level where number of dopants seems to be playing an important role on the etching process. Visible photoluminescence have been observed from all the porous samples possibly due to quantum confinement effect.
\end{abstract}
\maketitle
\section{Introduction}
\ \ \ \ Nanostructure/porous silicon(Si) attracted the attention of researchers worldwide first due to its visible light emitting properties exhibited at room temperature [1,2]  Si nano-structures (NSs) such as Si nano-crystals, porous Si, quantum dots, nano-wires (NWs) showed interesting properties [3,4] such as high density electronic states, enhanced thermoelectric properties, high surface to volume ratio etc. This provides a promising application potential in advanced electronic devices [5], optoelectronic devices [6,7], biological and chemical sensors [8–-10] etc. Various methods have been reported in literature for synthesis of Si NSs such as ion implantation technique [11–-14], electrochemical etching technique [15–-18] Laser Induced etching technique [19–-21], metal assisted chemical etching technique or metal induced etching (MIE) technique [22–-25] etc. A particular technique is chosen for fabrication of Si NSs based on suitability, availability and ease as each of these methods has its own advantages and disadvantages. Furthermore, mechanism of formation of Si NSs is different for different techniques. Out of these techniques, MIE is a simple and low-cost method which may be suitable for many practical usages like in large scale operations in industry [26]. In addition, using this method one can achieve precise positioning of aligned Si nano-wires as well as control of diameter, length, spacing and density [23] .In MIE, first NPs of noble metal (preferably Ag) are deposited on Si substrate (wafer) which acts as a catalyst to assist etching of Si in etching solution containing HF and $H_2O_2$. Part of Si substrate covered with AgNPs is etched much faster whereas little etching take place at the  uncovered surface, resulting in porous Si containing arrays of Si NWs depending on many factors like wafer orientation, doping level of Si substrate [24] etc. The process of etching of Si atoms from the substrate is same as that in electrochemical etching method. Details of exact mechanism of Si NWs formation is well known as reported in literature [22–-25].
Another important factor, which controls the formation of porous Si or Si NWs is the size of Ag NPs. It has been observed that a layer of Ag film with nominal thickness lead to fine pore structures in etching whereas columnar structures are formed with relatively thicker Ag film [27]. The Ag NPs act as etching sites from where the porosofication initiates and thus play a very important role in the etching process. These Ag NPs are converted to $Ag—$ ion during the etching process [25,28]. These $Ag—$ ions have a tendency of accepting electrons from the wafer and getting deposited as Ag NPs. Hence it is expected that this process may depend on initial doping level (or resistivity) of the starting Si wafer. It will be interesting to compare the porosification of different Si wafers with different doping levels. It is also observed that after MIE treatment, the substrate/sample usually looks black, which suggests a prominent antireflection property of the as-etched structures [24]. Thus, Si NWs formed using the above method may be used as a promising candidate for solar cells and photochemical solar cell devices.
Aim of the present paper is to compare the structures of porous Si  fabricated from two differently doped Si wafers. For direct comparison, two sets of samples have been fabricated from two n-type Si (100) wafers with different resistivities. For each set, three different etching times of 45, 60 \& 75 minutes have been used by keeping all other parameters (concentration of etching solution etc) constant. The samples have been investigated using SEM. It has been observed from the top view and cross- (X-) sectional view of all the samples that the higher resistivity wafers produce aligned Si nanowire arrays like structure after etching whereas relatively lower resistivity Si wafers produce an interconnected porous (cheese like) structure.
\section{Experimental details}
\ \ \ \ Commercially available Si (100) wafers with two different resistivities of 0.02 $\Omega$cm and 1-20 $\Omega$cm were used to fabricate porous structures in Si using MIE technique. All the wafers were cleaned in acetone and ethanol to remove impurities prior to starting the porosification process. The cleaned wafers were immersed in HF solution to remove any thin oxide layer formed at the Si surface. Then the wafers were dipped in solution containing 4.8 M HF \& 5 mM $AgNO_3$ for one minute at room temperature to deposit AgNPs. Three different Ag NPs deposited samples were then kept for etching in an etching solution containing 4.8 M HF and 0.5 M $H_2O_2$ for 45, 60 \& 75 minutes respectively for porosification. To remove extra Ag metal, the etched wafers were transferred in $HNO_3$ acid. After removal of the Ag metal, the samples were dipped into HF solution to remove any oxide layer induced by nitric acid used in the above step. List of all the samples has been provided in table-1. Surface morphology of all the samples have been characterized using Field-Emission SEM, supra55 Zeiss and Carl Zeiss in both plan-view and X-sectional geometries. The photoluminescence (PL) measurements have also been carried out with 325 nm laser excitation source using Dong Woo Optron 80K PL system at room temperature. 
\begin{center}
\begin{table}
\begin{tabular}{|c|c|c|}
  \hline                       
  Sample name & Wafer resistivity ($\Omega$ cm) & Etching time (Minutes) \\
  \hline
  H45 & 1-20 & 45 \\ 
  H60 & 1-20 & 60 \\
  H75 & 1-20 & 75 \\\hline
  L45 & 0.02 & 45 \\
  L60 & 0.02 & 60 \\
  L75 & 0.02 & 75 \\
  \hline
    
\end{tabular}
\caption{List of porous silicon samples investigated here }
\end{table}
\end{center}
\section{Results and Discussion}
\ \ \ \ Surface morphologies of all the samples (as listed in table-1) have been studied using SEM in top/plane view as well as X- sectional view to understand the etching of Si wafer for both the resistivity values.  SEM image corresponding to the porous Si samples fabricated from high resistivity (HR) wafers with resistivity 1-20$\Omega$cm have been shown in Fig. 1. Figures 1(a)–(c) show the top view whereas Fig. 1(d)-(f) show their corresponding X-sectional images. These images indicate that a dense and vertically aligned Si NWs arrays are formed similar to the already reported results [23-–25]. The formation of such SiNWs can be understood as follows. during the etching process, Ag NPs, present on the Si wafer, are partially oxidized by $H_2O_2$ to create a localized $Ag—$ cloud in the close proximity of the Ag NPs [24]. $Ag—$ ions quickly react with Si and eject electrons near the Ag/Si interface [28]. In this way, etching process starts locally around the Ag NPs and get trapped in the nano-pits created by very initial etching of the Si wafer, leading to continued etching in vertical direction. This leads to the formation of the SiNWs array.
The progress of etching of Si wafer with time can be seen by etching the wafers for longer in the same solution. It is expected that the SiNWs’ diameter and length (pore depth) should increase as a result of etching for longer duration [29]. In contrary, we observe a different result as shown in Fig. 1.  A careful analysis of the planar view of SEM images (Fig. 1(a)-(c)) shows that the diameter of Si NW arrays formed in sample H45 (Fig. 1(a)), is in the range of a few hundred nanometers, which increases to $\sim$ 2 $\mu$m for the case of H75 sample (Fig. 1(c)). Whereas, from the X-section SEM images we observe that the length of these Si NW arrays for the sample H45 (fig 1(d)) is around 60 $\mu$m which is decreased to a value upto $\sim$ 40 $\mu$m for H75 sample (fig 1(f)). Thus, we believe that the top of the Si NW arrays has broken and resulted in a blunt surface as the etching time increases. This is evident from the decrease of the length of Si NW arrays and increase of the diameter at higher etching time samples respectively. In contrary to our observation, Srivastava \emph{et al} [29]  have reported that with increasing etching time, length of the NWs increases from few nano-meters to several micrometers depending on the the etching time. The length of  SiNWs also increases linearly with etching time of 0 to 120 min [29]. But, in their case the initial Si wafers were p-type with (100) orientation. One can conclude from these observations that the optimum etching time for the formation vertically aligned Si NW arrays is $\sim$45 min in our case for n-type Si wafer with 1-20 $\Omega$cm resistivity. Further etching for longer duration may lead to breakage of the formed NW arrays.
\begin{figure}[h]
\begin{center}
\includegraphics[height=8cm]{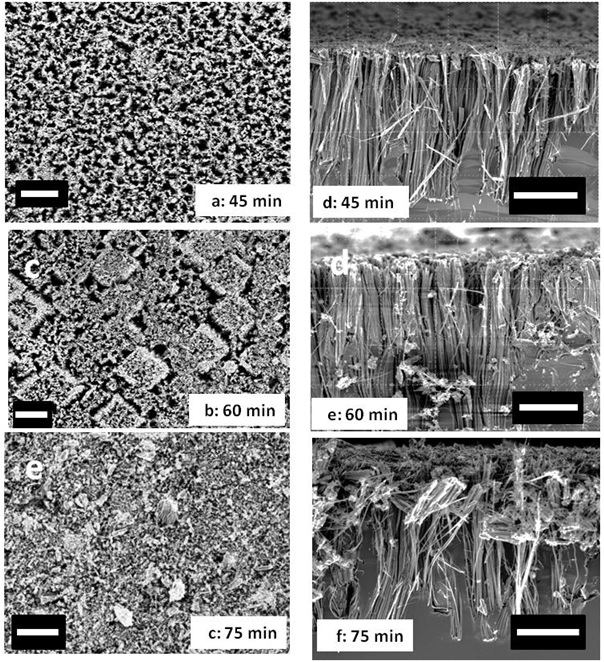}
\caption{SEM images of samples prepared for a given etching time using metal induced chemical etching of silicon wafer with resistivity of 1-20 $\Omega$cm in (a)-(c) top view and  and (d)-(f) X-sectional view. All the scale bars correspond to  20 $\mu$m}
\label{} 
\end{center}
\end{figure}
A schematic diagram showing step-by-step surface morphology reconstruction during MIE of HR Si has been shown in Fig. 2. Figure 2(a) shows the AgNPs deposited on Si wafers prior to the etching step. Formation of well-aligned pores as a result of MIE has been shown in Fig. 2(b). After further etching, the pore depth \& hence the SiNWs length increases upto a certain limit before it starts breaking because they cannot sustain the etching. Similar observation has been reported earlier for the case of porosification by laser induced etching [19]. 
To see the luminescence properties of these formed Si nano-wire arrays, photoluminescence (PL) measurements were carried out at room temperature using an excitation wavelength of 325 nm from He-Cd laser. The PL peak has been observed around 1.96 eV for all the samples as shown in Fig. 3. We observed only a little variation in PL peak positions as a function of etching time. As the etching time increases, from 45 minute to 60 minute the PL peak position shift to higher energy and further increase of the etching time reduced the PL energy for 75 minute etching. The PL peak position for the samples H45, H60 and H75 are 1.95, 1.97 and 1.96 eV respectively as shown in Fig. 3. At higher etching time of 75 min, due to broken nano-wire arrays the diameter of nano-wires is bigger as compare to that for 45 min case. This resulted in little red-shift in PL peak from 75 min etched samples as compared to 60 min etched sample. A room temperature visible PL from samples H45, H60, H75 indicates the presence of SiNWs having size in the range of few nanometers comparable to Bohr excitation radius. A PL peak position changing with etching time indicate the origin of PL to be due to quantum confinement [30].
\begin{figure}[h]
\begin{center}
\includegraphics[height=6cm]{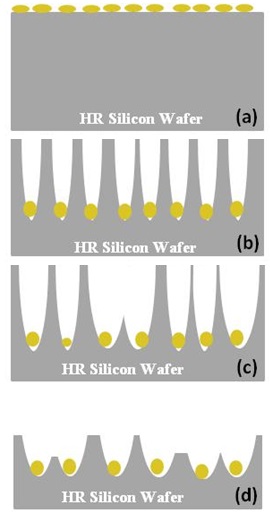}
\caption{Schematic diagram showing the etching process for high resistivity (1-20 $\Omega$cm ) silicon wafer.}
\label{} 
\end{center}
\end{figure}
After the discussion on the formation of Si NW arrays in case of high resistivity samples, porous Si samples fabricated from low resistivity Si wafer have also been studied. As discussed earlier, to study the effect of resistivity on Si porosification, we have repeated the etching process on comparatively low resistivity Si wafer (0.02 $\Omega$cm) keeping all the etching parameter fixed. Hereafter, these samples will be referred as LR samples. Fig. 4 shows SEM images (both planar and cross-sectional views) of the as-prepared LR samples at different etching times. One can clearly see from Fig. 4 that the pores are propagating through the intermediate Si channels making it a cheese like structure as the etching time increases. Interconnected pore formation has been observed for LR samples as compared to HR samples (discussed above) where well aligned SiNWs are observed.
The possible reason behind such difference in the porosification of Si wafer may be as follows. In a typical MIE process, the Ag NPs which act as catalyst, are pre-deposited on a clean Si wafer surface. The porous structures form in the close vicinity of these metal nanoparticles, as $H_2O_2$ alone cannot etch the Si faster. Ag NPs are partially oxidized by $H_2O_2$ to create a localized $Ag—$ ion in the close proximity of the silver particles. $Ag—$ ions can then quickly react with Si and take the electron from Si near Ag/Si interface and recovered into original Ag NPs. Hence, etching is localized around the Ag nanoparticle and further these nanoparticles are trapped in the pores created by them only. This lead to a vertical Si NW array formation as discussed in the previous section for HR Si case. In the case of LR samples, dopant concentration in the Si wafer is very high which resulted in the increase of the concentration of $Ag—$ ions. These $Ag—$ ions out-diffuse and nucleate on the side walls of Si pores formed  near a defective site (dopants present in Si wafer) or around the dopants [28]. By following the same etching process, new etching sites along the lateral direction of the SiNWs formed, in addition to the vertical etching path. The Ag NPs, nucleated on the pore walls, induce the pore formation in a direction which is not vertical. It is important here to mention that in addition to these non-vertical etching, the usual vertical pores are also formed. As the etching proceeds, the non-vertical porosofication started from one vertical pore reaches another vertical pore \& makes an interconnected pore pair. This kind of porosification at various random sites create several interconnected pores \& end up in the formation of cheese like porous Si as shown in Fig. 4. This explains the mechanism of formation of pores in lateral direction which lead to interconnected pores or a cheese like structure as observed in case of LR samples. This also explains why in lightly doped Si wafer (i.e. HR samples) only NW arrays are formed, as they have lesser carrier concentration /nucleation sites which give less probability for out-diffused $Ag—$ ions to nucleate on the side walls.
\begin{figure}[h]
\begin{center}
\includegraphics[height=6cm]{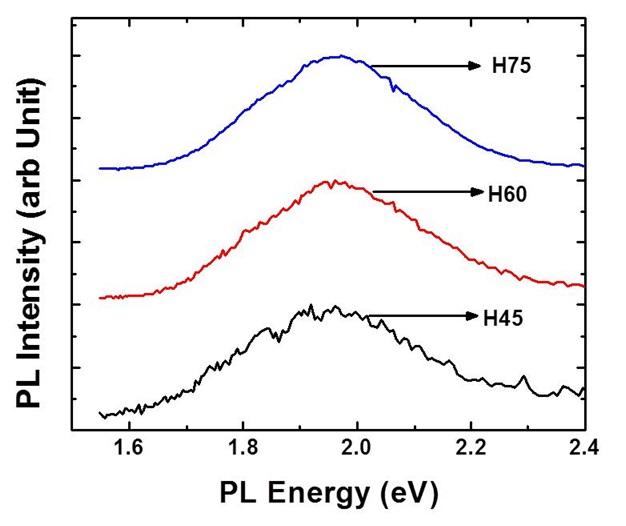}
\caption{Photoluminescence spectrum from High resistivity (1-20 $\Omega$cm ) MIE sample with etching time of (a) 45 minutes , (b) 60 minutes and (c) 75 minutes.}
\label{} 
\end{center}
\end{figure}
The effect of cheese-like structure or interconnected pores has also been seen in PL study of the LR samples. A broad PL peak around 1.91 eV has been observed from LR samples as shown in Fig. 5. On the other hand, for HR samples, the PL  peaks were observed at higher energy i.e. around 1.96 eV. The reason behind this difference  may be due to the fact that effective quantum confinement in the case of  HR is more as the NW arrays are not connected from each other and the effective diameter/size of NWs is small. In contrary if the pores are interconnected (as is the case for LR wafer), less confinement effect is expected resulting in NWs with bigger effective nanostructures sizes. Similar to HR samples, there is a slight variation in peak position and width with increasing etching time in LR samples as well.\begin{figure}[h]
\begin{center}
\includegraphics[height=8cm]{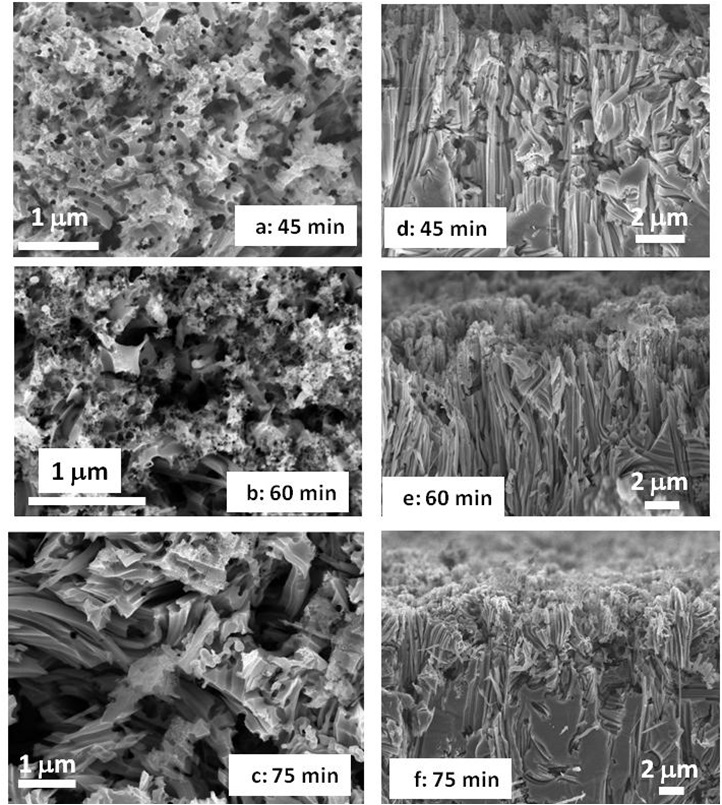}
\caption{SEM images of samples prepared for a given etching time using metal induced chemical etching of silicon wafer with resistivity of 0.02 $\Omega$cm in (a)-(c) top view and  and (d)-(f) X-sectional view. }
\label{} 
\end{center}
\end{figure} There is a red-shift in peak position with increasing etching time. Energy varies from 1.93 eV to 1.91 eV. This is very much understood as with increasing etching time, the probability of the pore formation in lateral direction increases which lead to increase in size of Si nano-structures leading to lower PL energy. The PL spectra corresponding to different etching time of LR samples has been shown in Fig. 5.
\begin{figure}[h]
\begin{center}
\includegraphics[height=6cm]{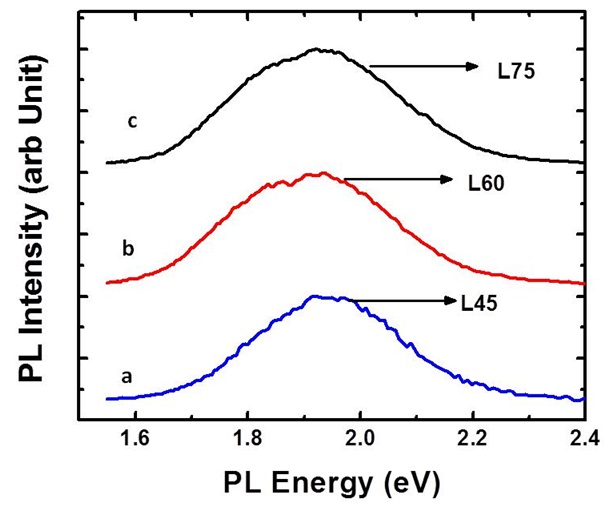}
\caption{Photoluminescence spectrum from Low resistivity (0.02 $\Omega$cm ) MIE sample with etching time of (a) 45 minutes , (b) 60 minutes and (c) 75 minutes.}
\label{} 
\end{center}
\end{figure} 
\section{Conclusions}
  
A comparative study of porous Si structures formed from two different Si wafers with different resistivities reveals that initial doping level of the starting Si wafer plays an important role on the pore formation. The comparison of surface morphologies and photoluminescence properties of porous Si fabricated using metal induced etching technique suggest that well aligned Si nanowires are formed from higher resistivity (1-20 $\Omega$cm) silicon wafer whereas interconnected pores or cheese like structures are formed from low resistivity Si (0.02 $\Omega$cm) wafers after MIE.  It was also observe that etching for more than 45 min in higher resistivity wafer lead to breakage of Si NWs formed as a result of chemical etching whereas in lower resistivity samples, at the same etching time, formation of highly connected pores leading to a cheese-like structures are formed. Visible photoluminescence has been observed from all the porous samples possibly due to quantum confinement effect. The extent of quantum confinement effect was found to be lesser in cheese like porous structure whereas relatively stronger quantum confinement effect was observed for well aligned non-connected pores.
 
\section*{Acknowledgements}
 Authors thank sophisticated instrumentation center (SIC), IIT Indore  for SEM \& PL measurements. One of the authors (GS) would like to acknowledge DST, Govt. of India, for the funding under DST Fast Track Scheme for Young Scientists, Project No. SR/FTP/PS-007/2012.
\end{document}